# Gravito-Electromagnetic Properties of Superconductors
## - A Brief Review -

C. J. de Matos[1], M. Tajmar[2]

June 20th, 2003

Starting from the generalised London equations, which include a gravitomagnetic term, the gravitational and the electromagnetic properties of superconductors are derived. A phenomenological synthesis of those properties is proposed.

**Table of Contents**

---

[1] Advanced Concepts and Studies Officer, ESA-HQ F-Paris, e-mail: clovis.de.matos@esa.int

[2] Principal Scientist, ARCS, A-Vienna, e-mail: martin.tajmar@arcs.ac.at

## I) Introduction

Starting from the generalised canonical momentum of Cooper pairs, which is proportional to the gradient of the phase of the wave function, and includes a magnetic and a gravitomagnetic vector potential, it is possible to generalise London's equations [Ross], and the fluxoid quantum condition law, such as to include gravitational effects. When we neglect gravity, the generalised London equations lead to the well-known electromagnetic properties of SCs. However when we consider simultaneously electromagnetic and gravitational interactions or gravitational interactions alone, we are led to predict modified electromagnetic properties and new gravitational properties for SCs, those are still poorly understood.

## II) Generalised London equations

When we regard the whole SuperConductor (SC) as being in a condensate state, we can describe it by a single condensate wave function, and the canonical momentum of Cooper pairs comes proportional to the gradient of the phase of their wave function. [Feynman]

$$\vec{\pi}_s = \hbar \nabla \theta \qquad (1)$$

For the case in which we include GravitoMagnetic (GM) phenomena [DeWitt] the canonical momentum contains an additional GM term proportional to the GM vector potential:

$$\vec{\pi}_s = m_p \vec{v}_s - \bar{q}_p \vec{A} + m_p \vec{A}_g \qquad (2)$$

Where $m_p = 2m$ is the mass of a Cooper pair, which is twice the free electron mass $m$, $\bar{q}_p = -2e$ is the electric charge of Cooper pairs, equal to twice the electron charge. $\vec{v}_s$ Is the Cooper pair classical cinematic velocity, $\vec{A}$ is the magnetic vector potential and $\vec{A}_g$ is the GravitoMagnetic (GM) vector potential.

## II-a) Second generalised London equation

From (1) we deduce that the rotational of Cooper pairs canonical momentum is always null

$$\nabla \times \vec{\pi}_s = 0 \tag{3}$$

Doing (2) into (3) we deduce we deduce the second generalised London equation

$$\boxed{\vec{v}_s = -\frac{e}{m}\vec{A} - \vec{A}_g} \tag{4}$$

Note that the gradient of the gravitational scalar potential never contributes to (4) since its Curl is zero.

## II-b) First generalised London equation

Doing (2) into (3) and taking the first derivative of (3) with respect to time, and substituting $\dot{\vec{A}}$ and $\dot{\vec{A}}_g$ by

$$\dot{\vec{A}} = -\nabla \phi_e - \vec{E} \tag{5}$$

$$\dot{\vec{A}}_g = -\nabla \phi_g - \vec{g} \tag{6}$$

Where $\phi_e$ and $\phi_g$ are respectively the electric and gravitational scalar potential. We obtain the first generalised London equation

$$\boxed{\dot{\vec{v}}_s = \frac{e}{m}\vec{E} + \vec{g}} \tag{7}$$

## III) Gravito-electromagnetic properties of SCs

From (4), (7) we can deduce all the electromagnetic and gravitational properties of SCs. These properties are revealed when we apply electromagnetic and / or gravitational fields to the SC or when the SC is set into motion. In this last case, as we will see below, there is a significant difference between the properties acquired by translation and those acquired by rotation.

*III-a) Generalised Meissner effect*

From (4) we can calculate the supercurrent generated in a SC when we apply simultaneously a magnetic and a GM field.

$$\nabla \times \vec{j}_s = -\frac{n_s e^2}{m}\vec{B} - n_s e \vec{B}_g \qquad (8)$$

Taking into account the 4$^{th}$ Maxwell equation and the 4$^{th}$ gravitational-Maxwell type equation, in stationary conditions:

$$\nabla \times \vec{B} = \mu_0 \vec{j}_s \qquad (9)$$

$$\nabla \times \vec{B}_g = -\mu_{0g} \vec{j}_{sm} \qquad (10)$$

Where $\vec{j}_{sm} = n_s m \vec{v}_s = \frac{m}{e}\vec{j}_s$ is the mass current density associated with the supercurrent density, and $\mu_{0g} = \frac{4\pi G}{c^2}$ is the gravitomagnetic permeability of vacuum. Doing (9) and (10) into the Curl of (8) we obtain the law of repartition of the supercurrent in the SC:

$$\nabla^2 \vec{j}_s - \frac{1}{\lambda_{GLPD}^2}\vec{j}_s = 0 \qquad (11)$$

With

$$\lambda_{GLPD} = \sqrt{\frac{m}{n_s\left(\mu_0 e^2 - \mu_{0g} m^2\right)}} \qquad (12)$$

From (11) and (12) we deduce that the supercurrent generated in a SC when we apply a magnetic and a GM field forms a layer, which will allow the magnetic field to penetrate in the material with an exponential attenuation. The characteristic decay length is $\lambda_{GLPD}$, called the Generalised London Penetration Depth (GLPD). The generalised London penetration depth is larger than the usual London Penetration Depth (LPD) $\lambda_{LPD} = \sqrt{\frac{1}{n_s \mu_0 e^2}}$ in the absence of a GM field: $\lambda_{GLPD} > \lambda_{LPD}$. Therefore a magnetic field will penetrate more in depth the SC when it is applied simultaneously with a GM field [Ross]

### *III-b) SCs do not shield gravitomagnetic and / or gravitational fields*

If we consider only a GM field in the rational of part III-a), the law of the repartition of the supercurrent in the SC is:

$$\nabla^2 \vec{j}_s - \frac{1}{\lambda_{GM}^2} \vec{j}_s = 0 \tag{13}$$

Where the GM London penetration depth: $\lambda_{GM} = i\sqrt{\dfrac{1}{\mu_{0g} n_s m}}$, is a complex number. That result is of course consistent with (12) above, when we neglect magnetism. Equation (13) shows that there is no GM Meissner effect, and that the concept of GM penetration depth has no physical meaning. Therefore a GM field will penetrate, the all bulk of a SC.

Taking the first derivative of (13) with respect to time, we obtain:

$$\nabla^2 \dot{\vec{v}}_s - \frac{1}{\lambda_{GM}^2} \dot{\vec{v}}_s = 0 \tag{14}$$

As we are only considering gravitational phenomena, we can substitute $\dot{\vec{v}}_s$ in (14), by (7) after setting $\vec{E} = 0$.

$$\nabla^2 \vec{g} - \frac{1}{\lambda_{GM}^2} \vec{g} = 0 \tag{15}$$

Since $\lambda_{GM}$ is a complex number, the lines of a gravitational field cannot be expelled from the SC; in other words, it is not possible to use SCs to shield gravity. (15) Has been obtained by M. Agop [Agop] but with a wrong GM penetration depth: $\lambda_{GM} = \sqrt{\dfrac{1}{\mu_{0g} n_s m}}$ [Agop].

Note that taking the Curl of (9) and (10) and using (8), having in attention $\vec{j}_{sm} = \frac{m}{e}\vec{j}_s$, we obtain the differential equations that describe how the magnetic and the GM fields penetrate the SC.

$$\nabla^2 \vec{B} - \frac{1}{\lambda_{LPD}^2}\vec{B} = \mu_0 n_s e \vec{B}_g \qquad (16)$$

$$\nabla^2 \vec{B}_g - \frac{1}{\lambda_{GM}^2}\vec{B}_g = -\mu_{0g} n_s e \vec{B} \qquad (17)$$

The conclusions we extract from these equations are similar to those that have already been presented above. Equations (16) and (17) have been obtained by N. Li [Li], but were not properly interpreted.

### III-c) *Generalised quantum fluxoid condition*

From (1) we deduce that the circulation of the generalised Canonical momentum of Cooper pairs around a closed path is quantised, since the value of the wave function must remain unchanged after a closed path, requiring:

$$\frac{1}{\hbar}\oint \vec{\pi}_s \cdot d\vec{l} = 2\pi n. \qquad (18)$$

Therefore:

$$\oint \vec{\pi}_s \cdot d\vec{l} = nh. \qquad (19)$$

Applying Stokes theorem to (9) we obtain the fluxoid quantum condition

$$n\frac{h}{2e} = \frac{m}{n_s e^2}\oiint \nabla \times \vec{j}_s \cdot d\vec{\sigma} + \oiint \vec{B} \cdot d\vec{\sigma} + \frac{m}{e}\oiint \vec{B}_g \cdot d\vec{\sigma} \qquad (20)$$

Where $n_s$ is the density of Cooper pairs, $\vec{j}_s = n_s e \vec{v}_s$ is the supercurrent density, $n$ is an integer and $h$ is the Planck's constant. Traditional interpretation of the flux quantisation relation (20) emphasize that what is physically measured in an experiment, is the observed magnetic flux $\phi_{m\,obs}$, such that:

$$\frac{m}{n_s e^2} \oiint \nabla \times \vec{j}_s \cdot d\vec{\sigma} + \phi_{m\,obs} = n\frac{h}{2e} - \oiint \vec{B} \cdot d\vec{\sigma} - \frac{m}{e} \oiint \vec{B}_g \cdot d\vec{\sigma} \qquad (21)$$

### *III-d) Supercurrents generated by GM fields*

Let us consider that we apply only a GM field to a SC. Computing the supercurrent, and the GM fluxes in (12) after setting $\vec{B} = 0$, and applying Stocke's theorem, it is possible to calculate the supercurrent generated in a simply connected SC when we apply only a GM field

$$\oint \vec{j}_s \cdot d\vec{l} = -n_s e B_g A \qquad (22)$$

Where $A = \pi R^2$ is the area defined by the contour of integration, that we can take arbitrarily as being defined by the average perimeter of the SC torus, with average radius $R$. That supercurrent will be present in the all bulk of the SC, since as shown above, the GM Meissner effect does not exist. Integrating (22) and taking the supercurrent density as the ratio between the electric current $I$ and the cross section of the torus $S = \pi r^2$, $j_s = \frac{I}{S}$, we can calculate the electric current generated through the application of a GM field to the SC torus:

$$I = \frac{n_s e R \pi r^2}{2} B_g \qquad (23)$$

The torus volume being $V = R \pi r^2$, and the Cooper pairs electric density being $\rho_s = n_s e$, (23) simplifies to

$$I = \frac{1}{2} Q_s B_g \qquad (24)$$

Where $Q_s = \rho_e V$ is the total electric charge associated with the Cooper pair's fluid. For lead there are about $3 \times 10^{28}$ electrons per $m^3$ contributing to the supercurrent, the GM field of the Earth on the equatorial plane is about $10^{-14} Rad/s$, if we assume that $R = 5\ cm$ and $r = 1\ cm$, the current will be $\approx 3.7 \times 10^{-10} A$ [DeWitt]. That current $I$ arises from an induced motion of the electrons in the bulk of the SC. If the applied GM field is caused by a rotating mass the motion of the electrons in the SC is in the same direction as the motion of the rotating mass, and represents another instance of the well-known drag effect produced by the Lense-Thirring field.

### III-e) Generalised London moment

Taking The Curl of (4) we deduce directly the generalised London moment [Tajmar 1]. In the absence of any external electromagnetic and gravitomagnetic field, a rotating SC ring with an angular velocity $\vec{\omega}$, rotated about its symmetry axis, generates in its interior a magnetic field $\vec{B}$, and a GM field $\vec{B}_g$,

$$\vec{B} = -\frac{2m}{e}\vec{\omega} - \frac{m}{e}\vec{B}_g \qquad (25)$$

The velocity of the Cooper pairs is equal to the velocity of the crystalline lattice of the SC, There are no magnetic and / or GM fields applied to the SC.

The London moment effect is understood as arising because the electrons near the surface "lag behind" when the body is put into rotation, and a surface current is generated. When the ions start moving the resulting electric current due to ionic motion generates a changing magnetic field, which in turn generates an electric field that makes the electrons follow suit. The relative magnitude between $\vec{B}$, $\vec{B}_g$ and $\vec{\omega}$ is an open question. As the magnetic and GM permeabilities do not enter into (25) the classical coupling between those fields might not be the classical one [Tajmar 3]. The value of $\vec{B}_g$ can be found by conjecturing that the presence of a GM field would resolve the Cooper pairs mass anomaly reported by Tate [Tajmar 2] [Tate]

$$\vec{B}_g = \frac{\Delta m}{m} 2\vec{\omega} \qquad (26)$$

Where $\Delta m$ is the Cooper pairs anomalous mass defect [Rae] and $m$ is the theoretical mass. We could call that phenomena the GM Barnett effect in quantum materials. For non-stationary conditions, doing (26) into the gravitational Faraday law ($\nabla \times \vec{g} = -\dot{\vec{B}}_g$), we predict the induction of non-Newtonian gravitational fields inside an angularly accelerated SCive ring of radius $R$.

$$\vec{g} = \frac{\Delta m}{m} R \dot{\omega} \hat{\theta} \qquad (27)$$

Where $\hat{\theta}$ is the azimuthal unit vector. This is currently under experimental assessment at ARC Seibersdorf Research.

### *III-f) Electric conductivity of SCs in a gravitational field*

The first generalised London equation (7) describes the resistanceless property of a SC, since in stationary conditions $\dot{\vec{v}}_s = 0$, and in the absence of gravity $\vec{g} = 0$, there is no electric field generated in the SC. Therefore it is not possible to construct the electromagnetic poynting vector $\vec{S} = \frac{1}{\mu_0} \vec{E} \times \vec{B}$, which integrated over the current carrier surface accounts for the power dissipated in the conductor.

However in presence of a gravitational field, and in stationary conditions there is a small electric field generated in the conductor, which is known as the Schiff-Barnhill effect [Schiff]

$$\vec{E} = -\frac{m}{e} \vec{g} \qquad (28)$$

In other words in the presence of a gravitational field, a SC is not a perfect conductor, but exhibits a tiny electrical resistance.

Let us investigate the consequences of the Schiff Barnhill effect on a closed current supported by a simply connected SC. We consider a closed squared SC torus with two of its sides parallel to the Earth gravitational field $\vec{g}_0 = g_0 \hat{z}$, on the surface of which circulates a closed electric current $i$ (see figure 1)

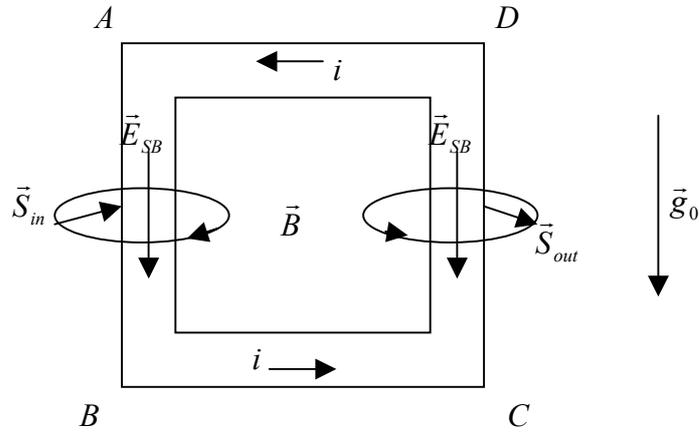

**Fig. 1**: SC squared ring in the Earth gravitational field.

Due to the Schiff-Barnhill effect the side $[AB]$ sees its temperature increasing, because it absorbs a power input:

$$P_{in} = \frac{m}{e} g B 2\pi r L \qquad (29)$$

$r$ being the radius of the cross section of the torus, and $L$ the length of its sides, $\vec{B}$ is the magnetic field generated by the closed SC current. On the other hand, the side $[CD]$ sees its temperature decreasing by the same amount as it increased for the side $[AB]$, because it releases a power output exactly opposite to the power input.

$$P_{out} = -P_{in} \qquad (30)$$

Therefore the thermodynamic Carnot engine we obtain with that configuration of the SC with respect to the Earth gravitational field, the Cooper pairs fluid being the working fluid, has null efficiency $\eta$. No work can be produced between two sources of heat at the same temperature

$$\eta = 1 - \frac{T_{AB}}{T_{CD}} = 0 \tag{31}$$

The SC will simply tend toward the thermal equilibrium without generating any work.

Witteborn and Fairbank demonstrated that the Schiff-Barnhill electric field is also present inside a SC cavity. We could envisage to locate a toroidal magnetic coil inside an hollow SC tube with coaxial axes (see figure 2), in order to release an output EM power superior to the electric power requested to maintain the magnetic field in the coil. The efficiency of such device, computed from the Poynting vector formed by the Schiff-Barnhill electric field and the magnetic field generated by the coil, is:

$$\eta = \frac{P_{out}}{P_{in}} = \frac{1}{U} \frac{m}{e} g_0 \frac{Nr^3}{R} \tag{32}$$

$N$ being the total number of turns of the coil, $r$ the radius of its section and $R$ the average radius of the torus, $P_{in} = Ui$, $i$ being the current circulating in the toroidal coil.

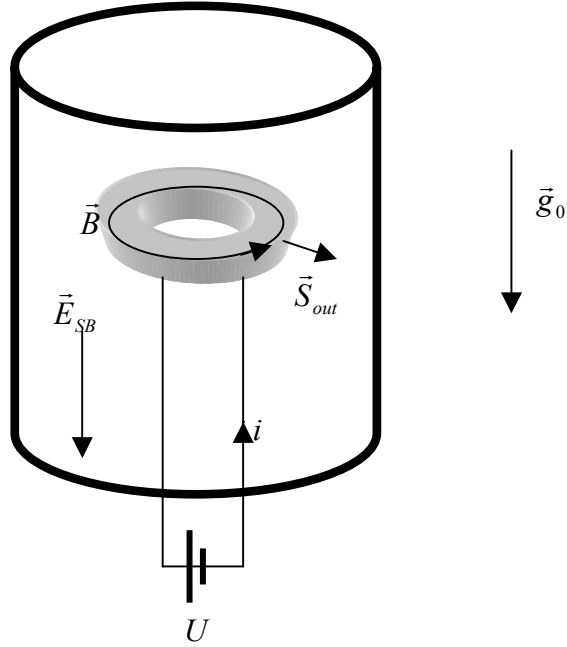

**Fig. 2**: Production of electric power out of the Schiff-Baarnhill effect with a classical toroidal magnetic coil inside a SC cavity.

We see that to have a significant output of energy produced in an efficient way, i.e. with $\eta > 1$ we need to be in a very strong gravitational field. In such regime the generalised London equations will assume a different form that will lead to a different expression for the Schiff-Barnhill effect making equation (28) obsolete [Ross]. We can conclude that the possibility to use the Schiff-Barnhill effect in SCs, in the Earth gravitational environment, for the aim of producing electromagnetic energy out of the energy stored directly in the Earth gravitational field ($\varepsilon_g = -\frac{1}{4\pi G} g_0^2 \approx -10^{10} J/m^3$), is not encouraging.

### III-g) Electrical fields in accelerated SCs

From (7) we see that a radial electric field will be present inside a SC rotating with angular velocity $\omega$.

$$\dot{\vec{v}}_s = \frac{v_s^2}{R}\hat{n} = \frac{e}{m}\vec{E} \tag{33}$$

This electric field generates an electric force on the electrons directed inwards, that compensates for the centrifugal force.

$$\vec{E}_n = \frac{m}{e}\omega R\,\hat{n} \tag{34}$$

From (7) we see also that if an angular acceleration $\dot{\omega}$ is communicated to the rotating SC, an azimutal electric field must be generated inside the SC, in order to compensate for the tangential acceleration.

$$\dot{\vec{v}}_s = R\dot{\omega}\,\hat{t} = \frac{e}{m}\vec{E} \tag{35}$$

$$\vec{E}_t = \frac{m}{e}R\dot{\omega}\,\hat{t} \tag{36}$$

Translational accelerations of the SC would lead to Schiff-Barnhill electric fields already mentioned in part III-f).

From (33) and (36) we see that the electrons move in core with the crystalline lattice of a SC when we rotate it. In other words, the principle that the net static force on a superconducting electron inside the body of a rotating SC must vanish; must always be valid. Therefore in rotating SC, in the absence of any external electromagnetic and gravitational fields, Coriolis and centrifugal forces experienced by the superelectrons must be compensated by corresponding electromagnetic and gravitational type lorentz forces [Rystephanic].

We would like also to stress that an electric field applied to a SC cavity cannot induce a field of acceleration in its interior (this would be the electric analogue of the generalised London moment effect, or the converse of the Schiff-Barnhill effect). This reveals a profound asymmetry between rotation and translation for macroscopic quantum materials as was already observed for the case of atomic quantum systems.

**IV) Conclusion**

In table 1 below we summarise the phenomenological relationships between electromagnetism, inertia and weak gravitational fields. Linearised relativity, inertia, and electromagnetism are characterised by a pair of fields, respectively:

- The gravitoelectric (or Newtonian gravitational field), and the gravitomagnetic field
- The acceleration and angular velocity fields
- The electric and the magnetic field

Each field is associated with a source:

- The gravitoelectric field is generated by the gravitational mass
- The gravitomagnetic field acts on the gravitational angular momentum
- The acceleration is felt by the inertial mass
- The angular velocity acts on the inertial angular momentum
- The electric field is created by the electric charge
- The magnetic field interact with the magnetic moment

The links between electromagnetism, inertia and gravity can be seen as physical induction phenomena, which allow to generate one field from another.

The bridges between inertia and linearised gravitation are well known:

- Acceleration relates with gravitoelectric fields via the weak principle of equivalence. A field of acceleration is equivalent locally to a gravitational field. This phenomena is encompassed in the weak

- equivalence principle, through the Einstein's elevator "gedanken" experiment.
- Angular velocity is related with the gravitomagnetic field through the gravitational Larmor theorem. This theorem establishes a formal equivalence between a rotating reference frame and a frame at rest in a gravitomagnetic field. It complements the Einstein elevator gedanken experiment to the extent that it makes it consistent with linearised general relativity [Mashhoon]. The GM Barnett effect in quantum materials if found, would form another link between those fields.
- Induction phenomena between gravitoelectric and gravitomagnetic fields are theoretically possible, through the Faraday gravitational type induction law.

The phenomena linking electromagnetism with inertia are:
- The Barnett effect, which consists in the magnetization of a material by its rotation.
- The converse effect exists and is call the Einstein-de Haas effect: the magnetization of a material generates its rotation.
- Faraday's law of induction between electric and magnetic fields allows for conversion between these two fields.

Looking at table 1 below it is apparent that the rates of conversion of inertial fields into gravitational fields do not depend on the properties of matter, they depend merely on the properties of vacuum (or spacetime). On the contrary the induction phenomena between electromagnetism and inertia are regulated by the properties of material media.

Somehow quantum materials establish a bridge between matter and vacuum (spacetime). The first generalised London equation "unifies" from the phenomenological point of view, the electric – the acceleration – and the gravitational fields. The second London equation "unifies": the magnetic, the angular velocity – and the gravitomagnetic field. In non-stationnary conditions the generalised London moment effect can be used to induce non-Newtonian gravitational fields. In contrast with the case of rotating normal materials, the magnetic and GM fields generated through rotation of SCs are not equivalent fields leading to an equivalent magnetisation (gravitomagnetisation respectively) of the material, they are instead real fields existing outside the material. Moreover these fields cannot be explained by the alignment of atoms and / or electron's spin, under the influence of a mechanical torque. Instead they are associated with the very rotation of the bulk of the SC crystalline lattice.

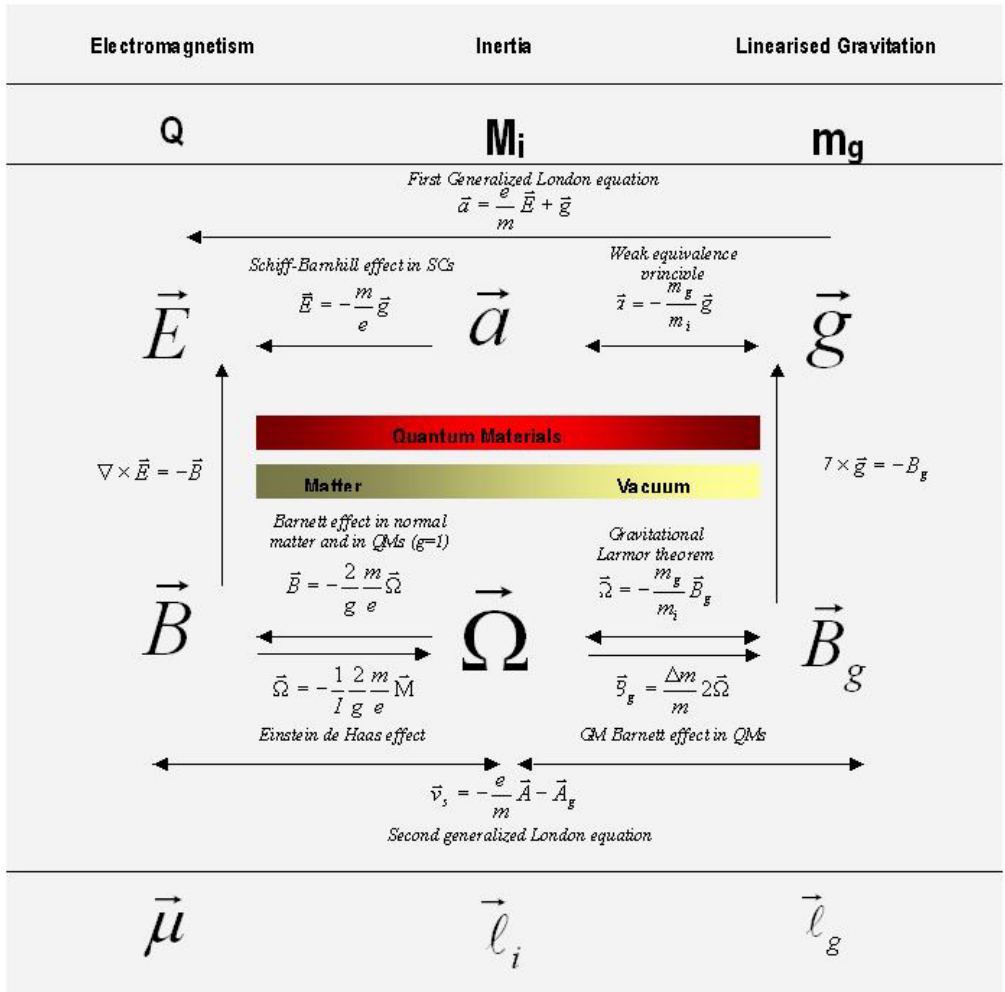

**Table 1**